\pdfoutput=1
\documentclass[preprint,12pt]{elsarticle}
\usepackage[english]{babel}
\usepackage{hyperref}
\usepackage{listings}
\usepackage{color}
\definecolor{grey}{rgb}{0.9,0.9,0.9}

\usepackage{graphicx}
\usepackage{lineno}

\journal{Computer Physics Communications}

\newcommand{\mykeywords}{ Electron density, projection, bilayer,
  cholesterol, lipid, phospholipid, membrane, trajectory, structural
  biology, graphical user interface, GUI, computational, molecular
  graphics %
}

\begin{document}

\begin{frontmatter}

\title{Computing 1-D atomic densities in macromolecular simulations: the Density Profile Tool for VMD}

\author{Toni Giorgino}
\ead{toni.giorgino@isib.cnr.it}
\address{Institute of Biomedical Engineering, National Research Council of Italy (ISIB-CNR)\\
Corso Stati Uniti 4, I-35127 Padua, Italy}

\begin{abstract}
  Molecular dynamics simulations have a prominent role in biophysics
  and drug discovery due to the atomistic information they provide on
  the structure, energetics and dynamics of biomolecules.  Specialized
  software packages are required to analyze simulated trajectories,
  either interactively or via scripts, to derive quantities of
  interest and provide insight for further experiments.  This paper
  presents the Density Profile Tool, a package that enhances the
  Visual Molecular Dynamics environment with the ability to
  interactively compute and visualize 1-D projections of various
  density functions of molecular models. We describe how the plugin is
  used to perform computations both via a graphical interface and
  programmatically.  Results are presented for realistic examples, 
  all-atom bilayer models, showing how mass and electron densities
  readily provide measurements such as membrane thickness, location of
  structural elements, and how they compare to X-ray diffraction
  experiments.
\end{abstract}

\begin{keyword}
  \mykeywords
\end{keyword}

\end{frontmatter}

{\bf Program summary}

\begin{small}
\noindent
{\em Manuscript Title:}                                       
 Computing 1-D atomic densities in macromolecular simulations: the Density Profile Tool for VMD \\
{\em Authors:}                                                
 Toni Giorgino \\
{\em Program Title:}                                          
 Density Profile Tool \\
{\em Journal Reference:}                                      \\
{\em Catalogue identifier:}                                   \\
{\em Licensing provisions:}                                   
 3-clause BSD Open Source. \\
{\em Programming language:}                                   
 TCL/TK. \\
{\em Computer:}                                               
 Any, with or without graphical display. \\
{\em Operating system:}                                       
 Linux/Unix, OSX, Windows. \\
{\em RAM:}                                               
  VMD should be able to hold the trajectory in memory. \\
{\em Number of processors used:}                              
 1 \\
{\em Keywords:} \mykeywords \\
{\em Classification:}                                         
  3 Biology and Molecular Biology, 23 Statistical Physics and Thermodynamics. \\
{\em External routines/libraries:}                          
  VMD (version 1.9 or higher).   \\
  {\em Nature of problem:} Compute and visualize one-dimensional
  density profiles of molecular dynamics trajectories in the VMD
  environment, either interactively or programmatically.
  \\
  {\em Solution method:} Density profiles are computed by binning the
  simulation space into slabs of finite thickness. A graphical user
  interface allows to choose the atomic property (number, mass,
  charge, electrons) and the details of the binning.
  \\
  {\em Restrictions:}
  The current version only supports orthorhombic cells. \\
  {\em Unusual features:} The Density Profile Tool is not a standalone
  program but a plug-in that enhances VMD's analysis features. \\
  {\em Running time:} A contemporary PC completes the analysis of  500 frames of the example system
  discussed in the paper (35,000 atoms)  in under 1 min. \\
\end{small}

\section{Introduction}
Molecular dynamics (MD) is a compute-intensive technique that
simulates the evolution of a system on the basis of interatomic forces
and Newton's laws of motion. MD has become a popular and successful
methodology to investigate the behavior of biomolecular systems at an
atomistic scale, because it provides structural information at
temporal and spatial resolutions much finer than those afforded by
most experimental
techniques~\cite{dror_biomolecular_2012,harvey_high-throughput_2012}.
Studies routinely use MD to probe the behavior of systems as diverse
as solvated globular proteins, biological
membranes~\cite{ash_computer_2004}, transmembrane
proteins~\cite{dror_identification_2009}, receptor
domains~\cite{ahmad_mechanism_2008,giorgino_visualizing_2012},
drug-protein interactions~\cite{buch_complete_2011,shan_how_2011},
etc. Predictivity is ensured by empirical force-fields which undergo
periodic refinements in order to faithfully reproduce structural,
energetic and kinetic observables~\cite{beauchamp_are_2012}.

\begin{figure}
  \centering
  \includegraphics[width=\textwidth]{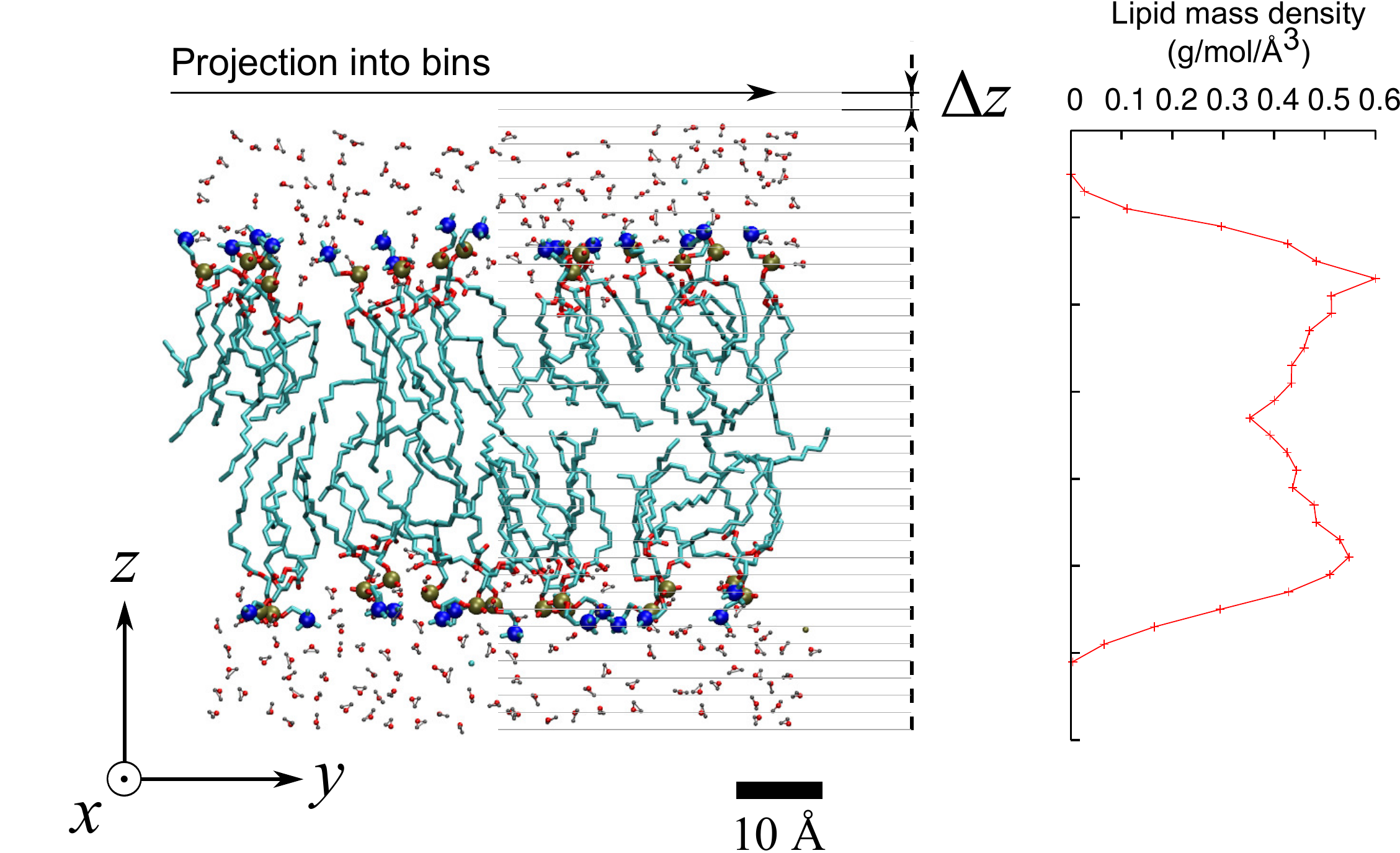}
  \caption{Left: computation of the density profile of a lipid bilayer
    whose normal is parallel to the $z$ axis. Profile resolution is
    $\Delta z = 2$ \AA. Right:  mass density profile computed for
    lipid molecules.}
  \label{fig:lipid}
\end{figure}

Density functions, such as number or mass densities, is one of the observables that can be derived from MD trajectories. For planar
systems, like biological membranes, it is convenient to project
density functions along the plane's normal to obtain 1-D profiles (Figure~\ref{fig:lipid}). The
profiles can be connected to experimental observables; for example,
the distance between peaks of lipid's phosphate groups is a direct
indicator of bilayer thicknesses, while  electron density profiles are
related to low angle diffraction and scattering
data~\cite{nagle_relations_1989}.

This paper describes the Density Profile tool, a software package that
computes one-dimensional density profiles of molecular systems. The
software is used from within the open-source Visual Molecular Dynamics
(VMD) environment, leveraging its extensive support for file formats,
interactive graphical display, atom selection syntax,
etc. \cite{Humphrey_Dalke_Schulten_1996} The plugin provides both a
graphical user interface (GUI), which makes it immediate to setup
computations, and a scripting interface.  GUI usage is generally
convenient to perform casual inspections, because no learning curve is
involved; on the other hand, the integration with VMD's scripting
environment, based on the TCL language~\cite{ousterhout_tcl_2009},
makes it possible to embed calculations in more complex protocols and
build on its results.

\subsection{Related work}

Other analysis packages contain command-line tools to perform density
profile computations similar to the one described here.  GROMACS'
distribution, for example, provides \verb+g_density+, a stand-alone
executable meant to be used from the command-line or in shell
scripts~\cite{hess_gromacs_2008}.  This approach has the advantage of
not being tied to a specific graphical or scripting environment;
however, it can be limiting in three respects: first, a GUI
 may be desirable for quick one-off computations; second,
binaries mostly require GROMACS-specific file formats and topologies;
finally, performing computations in shell scripts implies a
programming model in which  text files are used to store intermediate
results -- a model which is more cumbersome than
manipulating variables, the route afforded by
conventional programming languages like TCL or Python.

\section{Computational methods} %

Assuming point-like atoms (an assumption generally justified in
classical MD, see section~\ref{sec:limitations}), the algorithm to
compute density profiles is straightforward. For the sake of clarity,
here we assume to be interested in the mass density profile along the
$z$ axis, with a granularity of $\Delta z$.

\begin{figure}
  \centering
  \includegraphics[width=.6\textwidth]{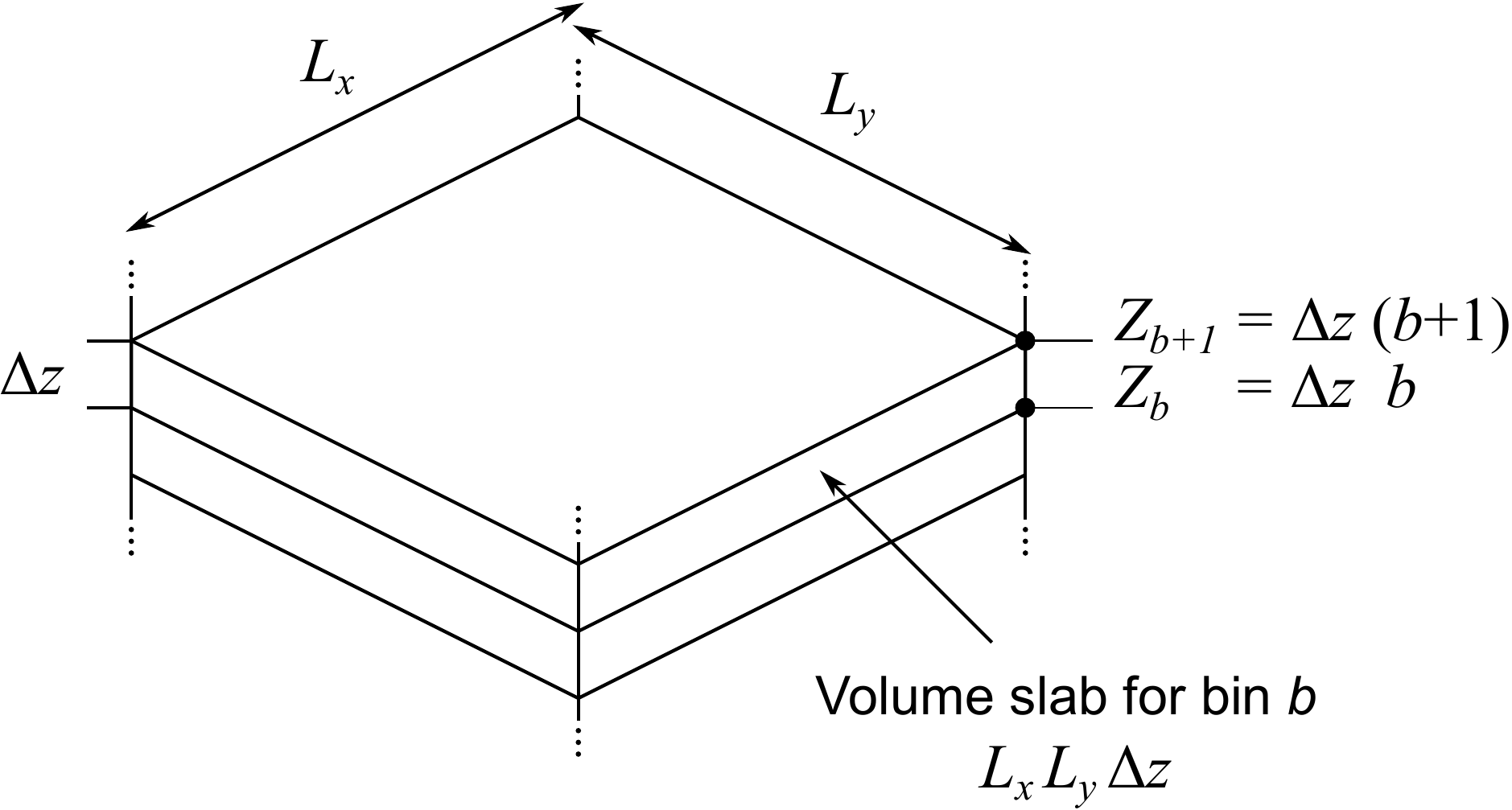}
  \caption{Arrangement of the volume slabs in a periodic cell.
    Density is being computed along the $z$ axis with a resolution of
    $\Delta z$. Bin $b$ (integer) contains atoms whose $z$ coordinate
    satisfies $b \le z / \Delta z < (b+1)$; each slab
    has a volume of $L_x L_y \Delta z$.}
  \label{fig:slabs}
\end{figure}

First, the space is divided in equally-sized slabs of thickness
$\Delta z$. Assuming orthorhombic periodic boundary conditions (PBC),
the slabs will extend in the $xy$ plane up to the boundaries of the
unit cell; otherwise, they can be considered infinitely large. Slabs are indexed
by a (relative) integer $b$; slab $b$ will include 
the spatial region $b \Delta
z \le z < (b+1) \Delta z$ (Figure~\ref{fig:slabs}). Defining the
indicator function $\delta_b(z)$ for slab $b$,
$$ \delta_b(z)=\left\{ 
  \begin{array}{rl}
    1 & \mbox{if $b \le z/\Delta z < b+1$} \\
    0 & \mbox{otherwise} 
  \end{array}
\right. $$
the value of the density function at bin $b$ is  obtained by binning
the chosen atomic property $p_i$, then normalizing by the slab volume:
\begin{equation}
  \label{eq:rho}
\rho_b = (L_x L_y \Delta z)^{-1}
\sum_{i \in \mathrm{atoms}} 
\delta_b(z_i) p_i 
\end{equation}
where $L_x$ and $L_y$ are the sides of the periodic cell, $z_i$ is the
$z$ coordinate of the center of atom $i$.  In the case of an infinite
system, it is possible to assume $L_x=L_y=1$ and obtain linear (as
opposed to volumetric) density values along $z$.

Depending on whether number,
mass, charge or electron number density is to be computed, the value
of $p_i$ is taken from different atomic attributes, according to
Table~\ref{tab:properties}.  These properties are
generally stored in the MD topology file associated with the
system.

\begin{table}
  \centering
  \begin{tabular}{cc}
    \hline \hline
    Density function    & $p =$ \\
    \hline
    Number              & 1 \\
    Mass                & $m_a$ \\
    Charge              & $q$ \\
    Electron number     & $Z-q$ \\
    Electron number (neutral) & $Z$ \\
    \hline \hline
  \end{tabular}
  \caption{Supported density function types and 
    the corresponding value to be used 
    for $p_i$ in Eq.~(\ref{eq:rho}). $m_a$ is the atomic
    mass, $q$ the  partial charge, $Z$ the atomic number. 
    In classical MD these properties can be
    derived from topology files, the atom's 
    name, and/or its type.}
  \label{tab:properties}
\end{table}

\section{Program description}

\begin{figure}
  \centering
  \includegraphics[width=.8\textwidth]{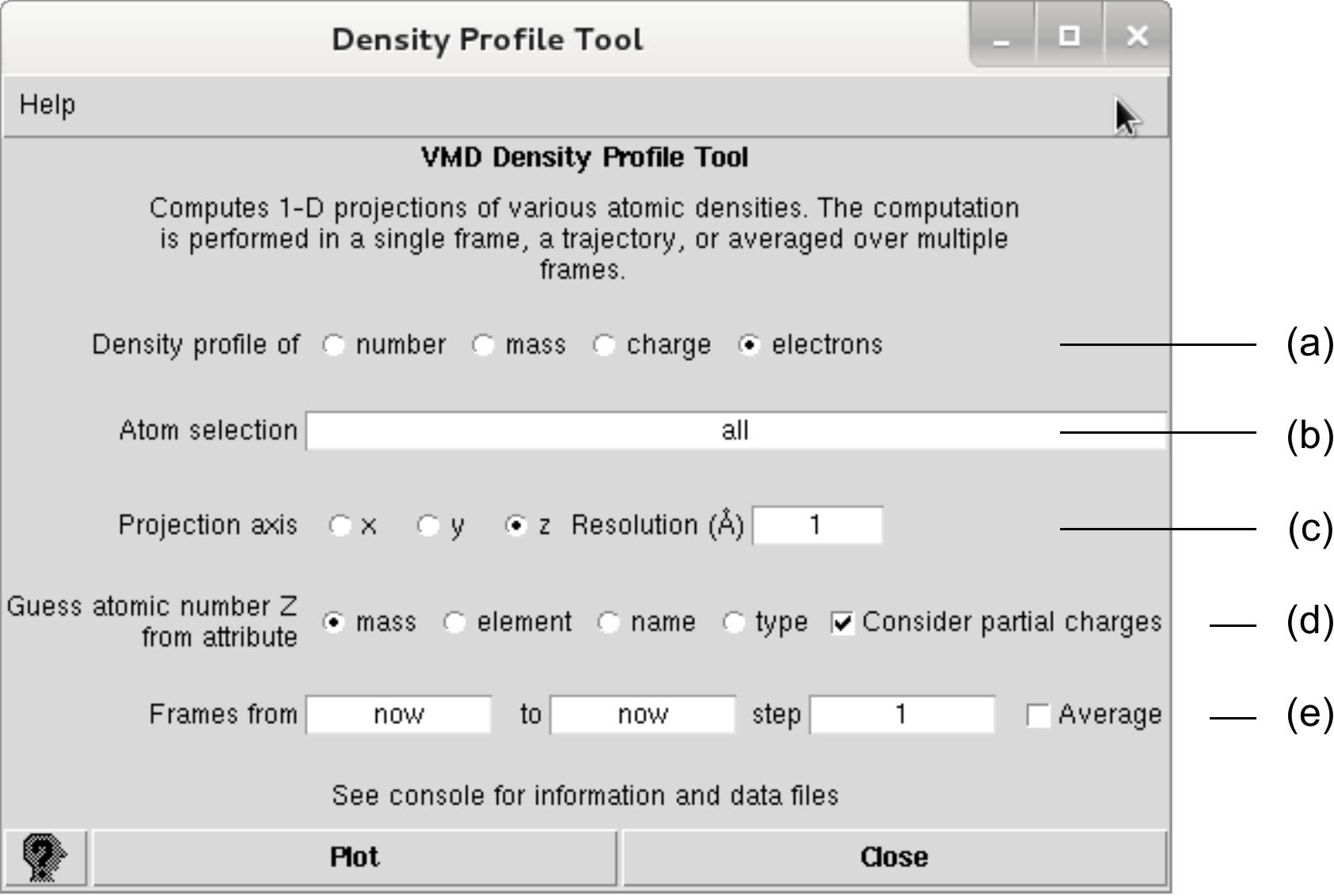}
  \caption{The plugin's graphical interface. Clicking the ``Plot''
    button starts the computation and shows the results in a plot
    window, with the option of exporting them as tabular data
    files. Labels are referenced in the main text.}
  \label{fig:gui}
\end{figure}

\subsection{Graphical user interface}

The most direct way to perform density profile calculations is to
load the trajectory to be analyzed in VMD and open
the plugin's GUI via the \emph{Extensions--Analysis} menu entry. A window will
appear (Figure~\ref{fig:gui}) and offer the chance to set the
analysis parameters,  with sensible defaults.

The GUI  is meant to be
self-explanatory. The user is able to select the property whose
density is to be computed (Figure~\ref{fig:gui}, a); the subset of the atoms to consider
(b); and the projection axis and bin size (c). Atom subsets are
specified through VMD's high-level atom selection syntax, which allows
complex queries on the base of geometry, atom types, structural
elements, and so on.

In case of electron density computations, we remark that topology
files generally do not store the required atomic number
$Z$. Therefore, an option (d) is provided to infer $Z$ from other
atomic properties, such as masses, names, or types (the former being 
the most reliable). The heuristics to obtain $Z$ are provided by the
TopoTools package~\cite{topotools}.
The \emph{consider partial charges} option includes or excludes
partial charges from the electron density.

Finally, a subset of the trajectory frames can be selected for
analysis setting an interval and stride (e). If the \emph{average}
option is enabled, average and standard deviations of densities over
the chosen trajectory interval will be reported; otherwise, the selected
frames are analyzed individually.

The analysis is started by clicking \emph{Plot} button, after which 
results will appear in a 
plot window. The plot's \emph{Export} menu allows to save the
graph itself or tabular ASCII data for further
processing.

\lstset{basicstyle=\footnotesize\ttfamily,columns=fullflexible,
  frame=single,backgroundcolor=\color{grey}, language=tcl}

\begin{figure}
  \centering
  \includegraphics[width=.6\textwidth]{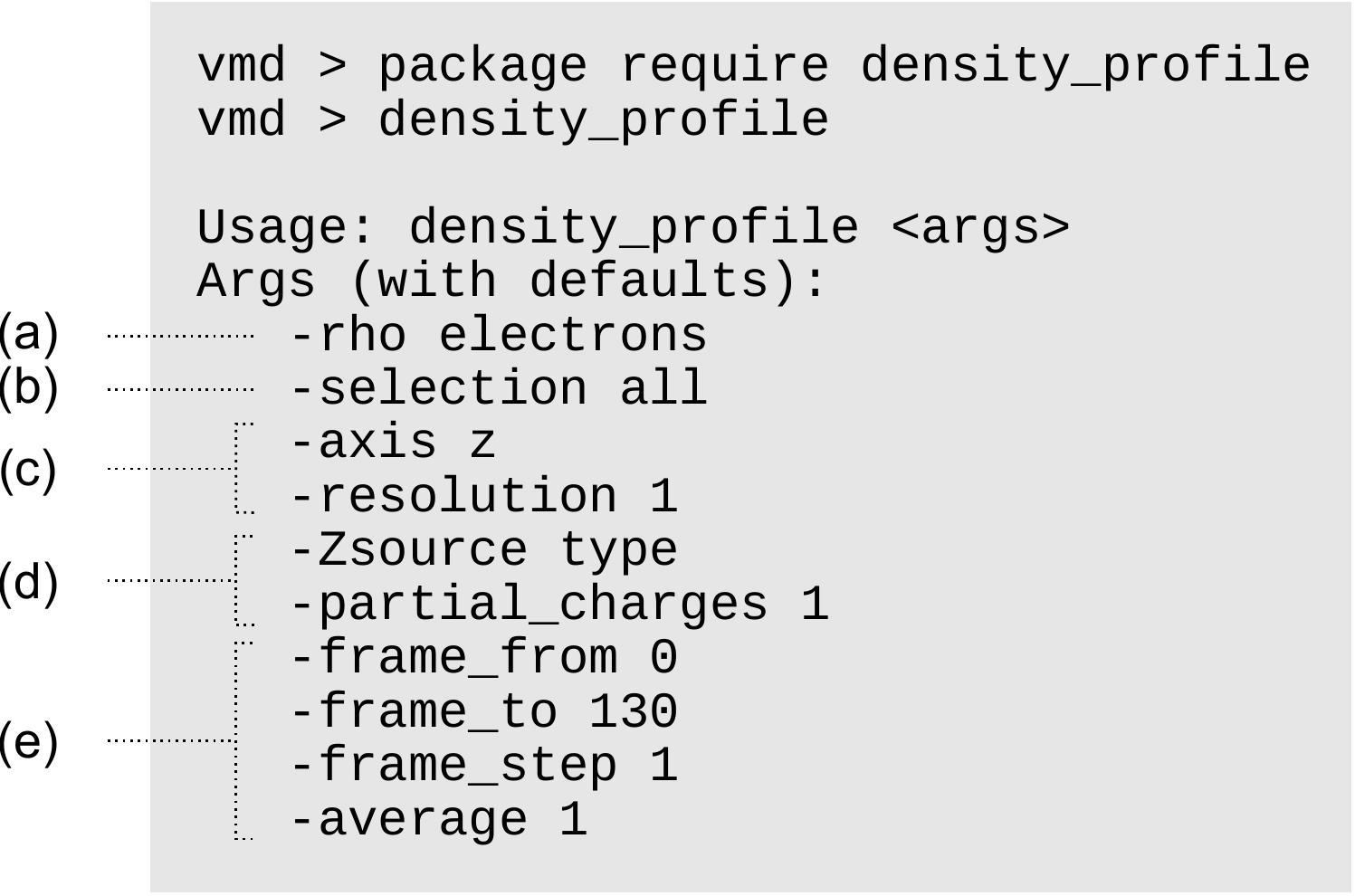}
  \caption{Command line interface for the density profile calculation
    function (with default values; 130 is the last loaded
    frame). Labels indicate the matching GUI controls in
    Figure~\ref{fig:gui}.}
  \label{fig:cliusage}
\end{figure}

\subsection{Command line interface}

Although the GUI makes it immediate to inspect
density profiles of the currently loaded trajectory, in practice the
need often arises to analyze series of simulations, or to use the results 
in a more complicated analysis protocol. For these
reasons, all of the plugin's functions can also be accessed through a
command-line interface, after loading the
package with the \verb+package require density_profile+ syntax. 

The tool's computations can be used in analysis programs written in
TCL (VMD's embedded language) via the
\verb+density_profile+ function. Calling the function
without arguments prints an usage summary on the console, with a list of  options and their
defaults (Figure~\ref{fig:cliusage}).  The options mirror the controls provided by the GUI;  for
example, the \verb+-rho+  argument  takes  one of the
\emph{number, mass, charge} or \emph{electrons} keywords; when
computing electron densities, \verb+-Zsource+ should be either
\emph{mass, element, name} or \emph{type}.

Invoking the function performs the calculation on the currently loaded
system and returns two TCL lists, each containing as many elements as
there are bins. The first list holds the density values of each bin;
the second list contains the corresponding lower coordinates, i.e.\
the locations of bin breaks.  If only one frame is selected, or the
``average'' option is enabled, the density values are scalar values
(one density value is returned per bin). If multiple frames are
selected (via the \verb+-frame_from+, \verb+_to+, and \verb+_step+
parameters), density values will themselves be lists, holding
results per bin and frame (see example in Section~\ref{sec:scripting}).

Provision of a double interface (mouse- and command-line driven) is
consistent with the rest of the VMD environment. The ability to use
plugins programmatically and to combine them into scripts is one of
the strengths of VMD, which enables the construction of arbitrarily
complex protocols.

\subsection{Units of measurement}

\begin{table}
  \centering
  \begin{tabular}{ccc}
    \hline \hline
                & \multicolumn{2}{c}{Unit cell} \\
     \cline{2-3}
    Function            & set          & not set \\
    \hline
    Number density      & \AA$^{-3}$    & \AA$^{-1}$  \\
    Mass density        & Da/\AA$^{3}$  & Da/\AA \\
    Charge density      & e/\AA$^{3}$   & e/\AA  \\
    Electron density    & \AA$^{-3}$    & \AA$^{-1}$  \\
    \hline \hline
  \end{tabular}
  \caption{Units of measurement of the computed
    densities. Unit \AA\ is $10^{-10}$ m; Da is one
    dalton, or unified atomic mass unit,  equal to 1 g/mol
    $\simeq 1.66 \times 10^{-27}$ Kg;  e $\simeq 1.60 \times 10^{-19}$ 
    coulomb is one elementary charge.}
  \label{tab:units}
\end{table}

The behaviour of the plugin differs depending on whether the size of
the periodic cell is available or not. In the former case, the
real-space volume of each slab is known, and the plugin computes the
expected \emph{volumetric} densities (Figure~\ref{fig:slabs}), 
normalized by cubic \AA ngstr\"om.
Following VMD's conventions, the units of measurements used are atoms
(number density), Dalton (mass), elementary charges (charge) or electron
number, as per Table~\ref{tab:units}.

If the volume of the unit cell is not known, or the simulation was
performed without PBC, the plugin will normalize densities with
respect to the bin width, thus reporting densities per unit of length.
Linear densities can then be normalized manually (which can be useful
e.g.\  to deal with non-orthorhombic cells).

\subsection{Limitations}\label{sec:limitations}

The algorithm considers point-like  atoms  instead of volumetric
distributions; the tool would therefore be inappropriate to compute
the density of isolated atoms with sub-\AA\ resolution (a setup
generally outside the scope of MD). The point-like atom
approximation is valid as long as the chosen resolution is coarser
than the average atomic radius, or the number of atoms in each bin is
large. The latter condition is commonly satisfied in the MD practice.

The current version of the package restricts simulations to those
performed in orthorhombic boundary conditions (either constant or
variable boxes).  It should be noted that these are the 
conditions used in most contemporary large-scale MD
simulations~\cite{shaw_atomic-level_2010,buch_high-throughput_2010}.

Finally, the plugin should provide correct results on coarse-grained and
united-atom models \cite{tozzini_minimalist_2010}, such as the Martini
force-field \cite{monticelli_martini_2008}, as well; this is due to
the fact that the coarse-grained topologies often contain the correct
atomic properties, with the exception of atomic numbers. For electron
density computations, ``united'' atomic numbers for the various atoms
types should be set manually (e.g. 10 for an united-atom water
molecule), because the heuristics to deduce atomic numbers is based 
on all-atom systems.

\section{Examples}

To illustrate the use of density profiles with a 
realistic use case, we
equilibrated a model bilayer of
1-palmitoyl-2-oleoyl-sn-glycero-3-phosphocholine (POPC) lipids in
presence and absence of cholesterol in a 4:1 ratio.  The hydrated
membrane systems were constructed with the CHARMM-GUI membrane builder
interface~\cite{jo_charmm-gui_2009,jo_charmm-gui_2008}, using a symmetric
composition for the two leaflets,  70 \AA\ initial box size, 15
\AA\ water buffer per side, and 0.15 M KCl ionic strength.
The generated systems, containing approximately 35,000 atoms each,
were parametrized with the CHARMM36 lipid
forcefield~\cite{klauda_update_2010} (cross-section shown
in Figure~\ref{fig:lipid}). After 1000 steps of
conjugate-gradient minimization, systems were simulated with the ACEMD software~\cite{harvey_acemd:_2009} for 100 ns  in the
constant-pressure (NPT) ensemble, holding fixed
the $x$-$y$ aspect ratio of the periodic box. 

\begin{figure}
  \centering
  \includegraphics[width=\textwidth,trim=0 2cm 0 0]{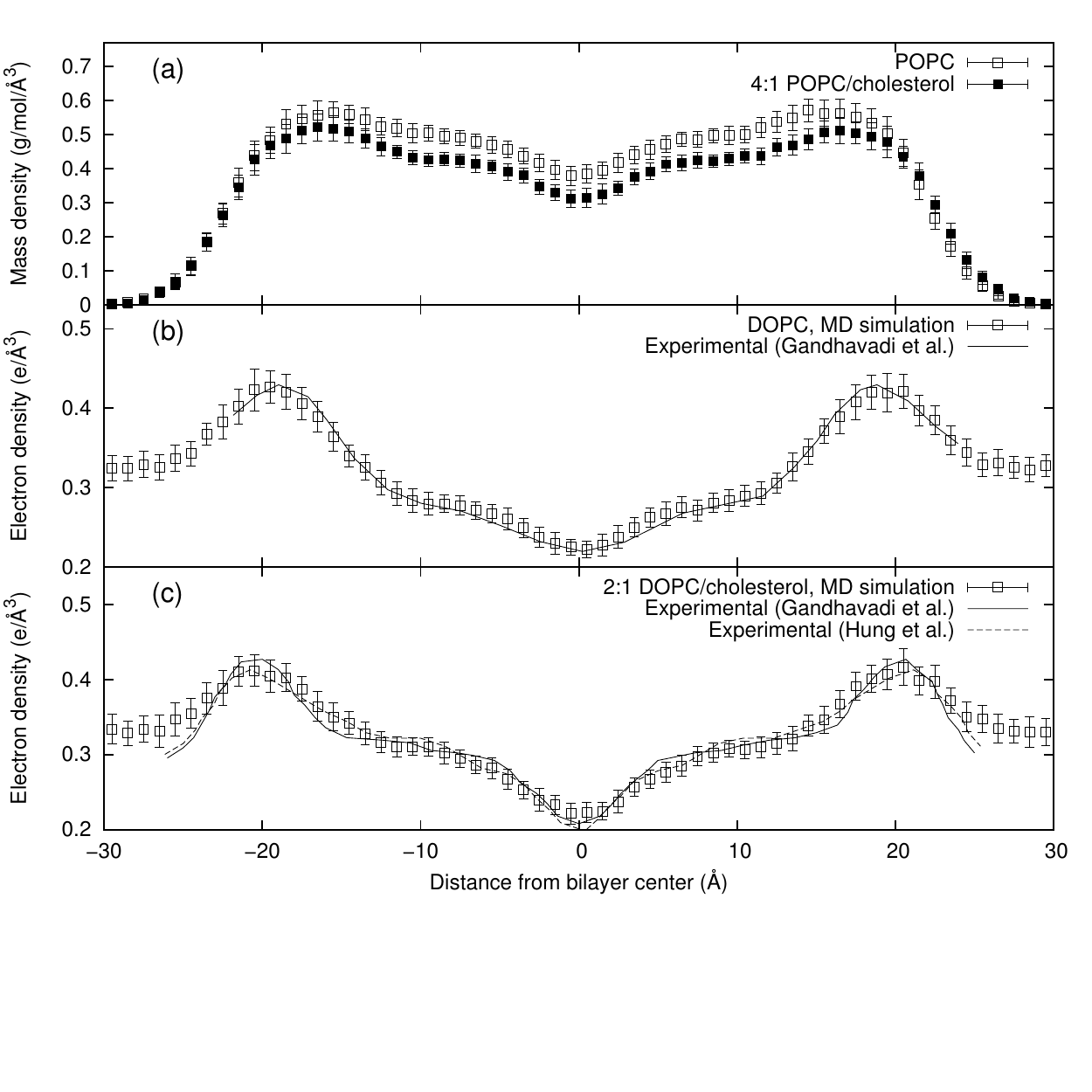}
  \includegraphics[width=\textwidth]{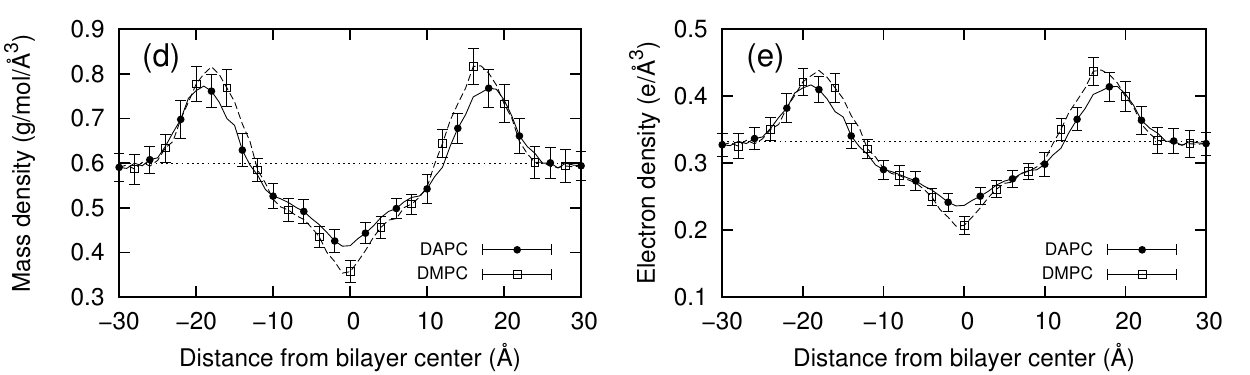}
  \caption{Density profiles in bilayer models. (a) Mass density
    distribution of POPC molecules in a pure-POPC bilayer (open
    squares) and a 4:1 POPC-cholesterol mixture (filled). (b) Electron
    density profiles of a pure-DOPC system: computed (squares) and
    experimentally-determined (solid
    line)~\cite[Fig.\ 5]{gandhavadi_structure_2002}. (c) Electron
    density profiles of a 2:1 DOPC-cholesterol system: computed
    (squares) and experimentally determined by Gandhavadi \emph{et
      al.}  \cite[Fig.\ 7]{gandhavadi_structure_2002} and Hung \emph{et
      al.}~\cite[Fig.\ 5]{hung_condensing_2007}. (d, e) Computed mass
    and electron density profiles for a polyunsaturated (DAPC) and
    a saturated (DMPC) fatty-acid phospholipid bilayer; dotted lines indicate
     densities of bulk water.}
  \label{fig:example}
\end{figure}

Density profiles were computed considering the last 50 ns of each
trajectory, using snapshots taken every 100 ps (500 frames per
simulation).  Figure~\ref{fig:example} (a) compares the final mass
distributions of POPC molecules of the two systems, shown  as
average and standard deviations along the analysis interval. As
expected, the bulky cholesterol molecules contribute to the membrane
surface area interposing between POPC molecules; therefore, POPC
volumetric mass density is decreased by approximately 15\%

Second, we computed the electron density of a pure
1,2-dioleoyl-sn-glycero-3-phosphocholine (DOPC) lipid bilayer,
prepared as above, and compared it with the experimental profile
determined through X-ray scattering by Gandhavadi 
\emph{et al.}\ \cite[Figure 5]{gandhavadi_structure_2002}. Figure~\ref{fig:example}
(b) shows that the experimental (solid line) and MD-derived profiles (points
with  standard deviations)  are in
excellent agreement.
Analogously, Figure~\ref{fig:example} (c) compares the 
computed profiles for  a
mixture of 2:1 DOPC-cholesterol  (points) with the experimental
results by Gandhavadi \emph{et al.}\ (solid line)~\cite[Figure
7]{gandhavadi_structure_2002}  and Hung \emph{et al.}\ 
(dashed line)~\cite[Figure 5]{hung_condensing_2007}; the agreement for
this heterogeneous bilayer is very good, as well.   Experimental
profiles, originally in arbitrary units, were scaled to fit.

As a last example, Figures~\ref{fig:example} (d,e) compare the mass
and electron density profiles of a polyunsaturated fatty acid
phospholipid bilayer (1,2-diarachidonoyl-glycero-3-phosphocholine,
DAPC) and a saturated one
(1,2-dimyristoyl-sn-glycero-3-phosphocholine, DMPC), both prepared as
above. The profiles include phospholipids together with solvent;
accordingly,  the density values outside of the bilayer region converge
towards the values expected for bulk water (dotted lines).

\subsection{Scripting}\label{sec:scripting}

For the sake of illustration, we perform the
same mass calculation of the above example, this time accessing the
results through the customary VMD-TCL language syntax:

\begin{lstlisting}
vmd > set dp [density_profile -rho mass -selection lipid
                        -frame_to 99 -frame_from 50 -frame_step 1]
vmd > puts [lindex $dp 0 3 5]     # Meaning dp(0,3,5)
vmd > puts [lindex $dp 1 3]        
\end{lstlisting}

The first line performs the mass density computation (without any
graphical interaction). The return value of the function call, a list
of two lists, is assigned to the variable \verb+dp+. The next command
prints the value of the density at bin 3 of the 6th selected frame
(trajectory frame 55). 
Analogously, the last line extracts the fourth element of the
second returned list, i.e.\ the lower coordinate of bin 3;
\verb+lindex+ is TCL's list index operator (indices are zero-based).

\section{Compatibility and distribution}

The density profile tool is written in TCL and is
platform-independent; it will therefore work on any of the platforms
supported by the  VMD system, notably Linux, OSX, Windows,
and numerous Unix variants \cite{Humphrey_Dalke_Schulten_1996}.  

The source is released under a 3-clause BSD license.  The latest
version of the package can be obtained from the URL
\url{multiscalelab.org/utilities/DensityProfileTool}. The archive also
contains cross-platform installation instructions.

\section{Acknowledgments}
Package development was started while at the Computational Biophysics
Laboratory (GRIB-IMIM-UPF) of the Universitat Pompeu Fabra with
partial support by the Ag\`encia de Gesti\'o\ d'Ajuts Universitaris i
de Recerca, Generalitat de Catalunya (2009 BP-B 00109).  These
institutions are gratefully acknowledged.

\bibliographystyle{model1-num-names}

\begin{thebibliography}{24}
\expandafter\ifx\csname natexlab\endcsname\relax\def\natexlab#1{#1}\fi
\providecommand{\bibinfo}[2]{#2}
\ifx\xfnm\relax \def\xfnm[#1]{\unskip,\space#1}\fi
%Type = Article
\bibitem[{Dror et~al.(2012)Dror, Dirks, Grossman, Xu, and
  Shaw}]{dror_biomolecular_2012}
\bibinfo{author}{R.~O. Dror}, \bibinfo{author}{R.~M. Dirks},
  \bibinfo{author}{J.~Grossman}, \bibinfo{author}{H.~Xu},
  \bibinfo{author}{D.~E. Shaw},
\newblock \bibinfo{title}{Biomolecular simulation: A computational microscope
  for molecular biology},
\newblock \bibinfo{journal}{Annual Review of Biophysics} \bibinfo{volume}{41}
  (\bibinfo{year}{2012}) \bibinfo{pages}{429--452}.
%Type = Article
\bibitem[{Harvey and De~Fabritiis(2012)}]{harvey_high-throughput_2012}
\bibinfo{author}{M.~J. Harvey}, \bibinfo{author}{G.~De~Fabritiis},
\newblock \bibinfo{title}{High-throughput molecular dynamics: the powerful new
  tool for drug discovery},
\newblock \bibinfo{journal}{Drug Discovery Today} \bibinfo{volume}{17}
  (\bibinfo{year}{2012}) \bibinfo{pages}{1059--1062}.
%Type = Article
\bibitem[{Ash et~al.(2004)Ash, Zlomislic, Oloo, and
  Tieleman}]{ash_computer_2004}
\bibinfo{author}{W.~L. Ash}, \bibinfo{author}{M.~R. Zlomislic},
  \bibinfo{author}{E.~O. Oloo}, \bibinfo{author}{D.~P. Tieleman},
\newblock \bibinfo{title}{Computer simulations of membrane proteins},
\newblock \bibinfo{journal}{Biochimica et Biophysica Acta ({BBA)} -
  Biomembranes} \bibinfo{volume}{1666} (\bibinfo{year}{2004})
  \bibinfo{pages}{158--189}.
%Type = Article
\bibitem[{Dror et~al.(2009)Dror, Arlow, Borhani, Jensen, Piana, and
  Shaw}]{dror_identification_2009}
\bibinfo{author}{R.~O. Dror}, \bibinfo{author}{D.~H. Arlow},
  \bibinfo{author}{D.~W. Borhani}, \bibinfo{author}{M.~O. Jensen},
  \bibinfo{author}{S.~Piana}, \bibinfo{author}{D.~E. Shaw},
\newblock \bibinfo{title}{Identification of two distinct inactive conformations
  of the beta2-adrenergic receptor reconciles structural and biochemical
  observations},
\newblock \bibinfo{journal}{Proceedings of the National Academy of Sciences of
  the United States of America} \bibinfo{volume}{106} (\bibinfo{year}{2009})
  \bibinfo{pages}{4689--4694}.
%Type = Article
\bibitem[{Ahmad et~al.(2008)Ahmad, Gu, and Helms}]{ahmad_mechanism_2008}
\bibinfo{author}{M.~Ahmad}, \bibinfo{author}{W.~Gu},
  \bibinfo{author}{V.~Helms},
\newblock \bibinfo{title}{Mechanism of fast peptide recognition by {SH3}
  domains},
\newblock \bibinfo{journal}{Angewandte Chemie (International Ed. in English)}
  \bibinfo{volume}{47} (\bibinfo{year}{2008}) \bibinfo{pages}{7626--7630}.
%Type = Article
\bibitem[{Giorgino et~al.(2012)Giorgino, Buch, and
  De~Fabritiis}]{giorgino_visualizing_2012}
\bibinfo{author}{T.~Giorgino}, \bibinfo{author}{I.~Buch},
  \bibinfo{author}{G.~De~Fabritiis},
\newblock \bibinfo{title}{Visualizing the induced binding of
  {SH2-Phosphopeptide}},
\newblock \bibinfo{journal}{J. Chem. Theory Comput.} \bibinfo{volume}{8}
  (\bibinfo{year}{2012}) \bibinfo{pages}{1171--1175}.
%Type = Article
\bibitem[{Buch et~al.(2011)Buch, Giorgino, and
  De~Fabritiis}]{buch_complete_2011}
\bibinfo{author}{I.~Buch}, \bibinfo{author}{T.~Giorgino},
  \bibinfo{author}{G.~De~Fabritiis},
\newblock \bibinfo{title}{Complete reconstruction of an enzyme-inhibitor
  binding process by molecular dynamics simulations},
\newblock \bibinfo{journal}{Proceedings of the National Academy of Sciences}
  \bibinfo{volume}{108} (\bibinfo{year}{2011}) \bibinfo{pages}{10184--10189}.
%Type = Article
\bibitem[{Shan et~al.(2011)Shan, Kim, Eastwood, Dror, Seeliger, and
  Shaw}]{shan_how_2011}
\bibinfo{author}{Y.~Shan}, \bibinfo{author}{E.~T. Kim}, \bibinfo{author}{M.~P.
  Eastwood}, \bibinfo{author}{R.~O. Dror}, \bibinfo{author}{M.~A. Seeliger},
  \bibinfo{author}{D.~E. Shaw},
\newblock \bibinfo{title}{How does a drug molecule find its target binding
  site?},
\newblock \bibinfo{journal}{Journal of the American Chemical Society}
  \bibinfo{volume}{133} (\bibinfo{year}{2011}) \bibinfo{pages}{9181--9183}.
%Type = Article
\bibitem[{Beauchamp et~al.(2012)Beauchamp, Lin, Das, and
  Pande}]{beauchamp_are_2012}
\bibinfo{author}{K.~A. Beauchamp}, \bibinfo{author}{Y.-S. Lin},
  \bibinfo{author}{R.~Das}, \bibinfo{author}{V.~S. Pande},
\newblock \bibinfo{title}{Are protein force fields getting better? a systematic
  benchmark on 524 diverse {NMR} measurements},
\newblock \bibinfo{journal}{Journal of Chemical Theory and Computation}
  \bibinfo{volume}{8} (\bibinfo{year}{2012}) \bibinfo{pages}{1409--1414}.
%Type = Article
\bibitem[{Nagle and Wiener(1989)}]{nagle_relations_1989}
\bibinfo{author}{J.~F. Nagle}, \bibinfo{author}{M.~C. Wiener},
\newblock \bibinfo{title}{Relations for lipid bilayers. connection of electron
  density profiles to other structural quantities.},
\newblock \bibinfo{journal}{Biophysical Journal} \bibinfo{volume}{55}
  (\bibinfo{year}{1989}) \bibinfo{pages}{309--313}.
%Type = Article
\bibitem[{Humphrey et~al.(1996)Humphrey, Dalke, and
  Schulten}]{Humphrey_Dalke_Schulten_1996}
\bibinfo{author}{W.~Humphrey}, \bibinfo{author}{A.~Dalke},
  \bibinfo{author}{K.~Schulten},
\newblock \bibinfo{title}{{VMD}: visual molecular dynamics},
\newblock \bibinfo{journal}{J Mol Graph} \bibinfo{volume}{14}
  (\bibinfo{year}{1996}) \bibinfo{pages}{33--38}.
%Type = Book
\bibitem[{Ousterhout and Jones(2009)}]{ousterhout_tcl_2009}
\bibinfo{author}{J.~K. Ousterhout}, \bibinfo{author}{K.~Jones},
  \bibinfo{title}{Tcl and the Tk Toolkit}, \bibinfo{publisher}{Addison-Wesley
  Professional}, \bibinfo{edition}{2nd} edition, \bibinfo{year}{2009}.
%Type = Article
\bibitem[{Hess et~al.(2008)Hess, Kutzner, van~der Spoel, and
  Lindahl}]{hess_gromacs_2008}
\bibinfo{author}{B.~Hess}, \bibinfo{author}{C.~Kutzner},
  \bibinfo{author}{D.~van~der Spoel}, \bibinfo{author}{E.~Lindahl},
\newblock \bibinfo{title}{{GROMACS} 4: Algorithms for highly efficient,
  load-balanced, and scalable molecular simulation},
\newblock \bibinfo{journal}{Journal of Chemical Theory and Computation}
  \bibinfo{volume}{4} (\bibinfo{year}{2008}) \bibinfo{pages}{435--447}.
%Type = Manual
\bibitem[{Kohlmeyer(2012)}]{topotools}
\bibinfo{author}{A.~Kohlmeyer}, \bibinfo{title}{TopoTools Plugin, Version 1.2},
  \bibinfo{year}{Accessed 7 May 2012}.
  \bibinfo{note}{\url{http://www.ks.uiuc.edu/Research/vmd/plugins/topotools/}}.
%Type = Article
\bibitem[{Shaw et~al.(2010)Shaw, Maragakis, Lindorff-Larsen, Piana, Dror,
  Eastwood, Bank, Jumper, Salmon, Shan, and Wriggers}]{shaw_atomic-level_2010}
\bibinfo{author}{D.~E. Shaw}, \bibinfo{author}{P.~Maragakis},
  \bibinfo{author}{K.~Lindorff-Larsen}, \bibinfo{author}{S.~Piana},
  \bibinfo{author}{R.~O. Dror}, \bibinfo{author}{M.~P. Eastwood},
  \bibinfo{author}{J.~A. Bank}, \bibinfo{author}{J.~M. Jumper},
  \bibinfo{author}{J.~K. Salmon}, \bibinfo{author}{Y.~Shan},
  \bibinfo{author}{W.~Wriggers},
\newblock \bibinfo{title}{Atomic-level characterization of the structural
  dynamics of proteins},
\newblock \bibinfo{journal}{Science} \bibinfo{volume}{330}
  (\bibinfo{year}{2010}) \bibinfo{pages}{341 --346}.
%Type = Article
\bibitem[{Buch et~al.(2010)Buch, Harvey, Giorgino, Anderson, and
  De~Fabritiis}]{buch_high-throughput_2010}
\bibinfo{author}{I.~Buch}, \bibinfo{author}{M.~J. Harvey},
  \bibinfo{author}{T.~Giorgino}, \bibinfo{author}{D.~P. Anderson},
  \bibinfo{author}{G.~De~Fabritiis},
\newblock \bibinfo{title}{High-throughput all-atom molecular dynamics
  simulations using distributed computing},
\newblock \bibinfo{journal}{Journal of Chemical Information and Modeling}
  \bibinfo{volume}{50} (\bibinfo{year}{2010}) \bibinfo{pages}{397--403}.
%Type = Article
\bibitem[{Tozzini(2010)}]{tozzini_minimalist_2010}
\bibinfo{author}{V.~Tozzini},
\newblock \bibinfo{title}{Minimalist models for proteins: a comparative
  analysis},
\newblock \bibinfo{journal}{Quarterly Reviews of Biophysics}
  \bibinfo{volume}{43} (\bibinfo{year}{2010}) \bibinfo{pages}{333--371}.
%Type = Article
\bibitem[{Monticelli et~al.(2008)Monticelli, Kandasamy, Periole, Larson,
  Tieleman, and Marrink}]{monticelli_martini_2008}
\bibinfo{author}{L.~Monticelli}, \bibinfo{author}{S.~K. Kandasamy},
  \bibinfo{author}{X.~Periole}, \bibinfo{author}{R.~G. Larson},
  \bibinfo{author}{D.~P. Tieleman}, \bibinfo{author}{S.-J. Marrink},
\newblock \bibinfo{title}{The {MARTINI} coarse-grained force field: Extension
  to proteins},
\newblock \bibinfo{journal}{Journal of Chemical Theory and Computation}
  \bibinfo{volume}{4} (\bibinfo{year}{2008}) \bibinfo{pages}{819--834}.
%Type = Article
\bibitem[{Jo et~al.(2009)Jo, Lim, Klauda, and Im}]{jo_charmm-gui_2009}
\bibinfo{author}{S.~Jo}, \bibinfo{author}{J.~B. Lim}, \bibinfo{author}{J.~B.
  Klauda}, \bibinfo{author}{W.~Im},
\newblock \bibinfo{title}{{CHARMM-GUI} membrane builder for mixed bilayers and
  its application to yeast membranes},
\newblock \bibinfo{journal}{Biophysical Journal} \bibinfo{volume}{97}
  (\bibinfo{year}{2009}) \bibinfo{pages}{50--58}.
%Type = Article
\bibitem[{Jo et~al.(2008)Jo, Kim, Iyer, and Im}]{jo_charmm-gui_2008}
\bibinfo{author}{S.~Jo}, \bibinfo{author}{T.~Kim}, \bibinfo{author}{V.~G.
  Iyer}, \bibinfo{author}{W.~Im},
\newblock \bibinfo{title}{{CHARMM-GUI:} a web-based graphical user interface
  for {CHARMM}},
\newblock \bibinfo{journal}{Journal of Computational Chemistry}
  \bibinfo{volume}{29} (\bibinfo{year}{2008}) \bibinfo{pages}{1859–1865}.
%Type = Article
\bibitem[{Klauda et~al.(2010)Klauda, Venable, Freites, {O’Connor}, Tobias,
  Mondragon-Ramirez, Vorobyov, {MacKerell}, and Pastor}]{klauda_update_2010}
\bibinfo{author}{J.~B. Klauda}, \bibinfo{author}{R.~M. Venable},
  \bibinfo{author}{J.~A. Freites}, \bibinfo{author}{J.~W. {O’Connor}},
  \bibinfo{author}{D.~J. Tobias}, \bibinfo{author}{C.~Mondragon-Ramirez},
  \bibinfo{author}{I.~Vorobyov}, \bibinfo{author}{A.~D. {MacKerell}},
  \bibinfo{author}{R.~W. Pastor},
\newblock \bibinfo{title}{Update of the {CHARMM} all-atom additive force field
  for lipids: Validation on six lipid types},
\newblock \bibinfo{journal}{The Journal of Physical Chemistry B}
  \bibinfo{volume}{114} (\bibinfo{year}{2010}) \bibinfo{pages}{7830--7843}.
%Type = Article
\bibitem[{Harvey et~al.(2009)Harvey, Giupponi, and
  Fabritiis}]{harvey_acemd:_2009}
\bibinfo{author}{M.~J. Harvey}, \bibinfo{author}{G.~Giupponi},
  \bibinfo{author}{G.~D. Fabritiis},
\newblock \bibinfo{title}{{ACEMD:} accelerating biomolecular dynamics in the
  microsecond time scale},
\newblock \bibinfo{journal}{Journal of Chemical Theory and Computation}
  \bibinfo{volume}{5} (\bibinfo{year}{2009}) \bibinfo{pages}{1632--1639}.
%Type = Article
\bibitem[{Gandhavadi et~al.(2002)Gandhavadi, Allende, Vidal, Simon, and
  {McIntosh}}]{gandhavadi_structure_2002}
\bibinfo{author}{M.~Gandhavadi}, \bibinfo{author}{D.~Allende},
  \bibinfo{author}{A.~Vidal}, \bibinfo{author}{S.~Simon},
  \bibinfo{author}{T.~{McIntosh}},
\newblock \bibinfo{title}{Structure, composition, and peptide binding
  properties of detergent soluble bilayers and detergent resistant rafts},
\newblock \bibinfo{journal}{Biophysical Journal} \bibinfo{volume}{82}
  (\bibinfo{year}{2002}) \bibinfo{pages}{1469--1482}.
%Type = Article
\bibitem[{Hung et~al.(2007)Hung, Lee, Chen, and Huang}]{hung_condensing_2007}
\bibinfo{author}{W.-C. Hung}, \bibinfo{author}{M.-T. Lee},
  \bibinfo{author}{F.-Y. Chen}, \bibinfo{author}{H.~W. Huang},
\newblock \bibinfo{title}{The condensing effect of cholesterol in lipid
  bilayers},
\newblock \bibinfo{journal}{Biophysical Journal} \bibinfo{volume}{92}
  (\bibinfo{year}{2007}) \bibinfo{pages}{3960--3967}.

\end{thebibliography}

\end{document}